# ARTICLE

# Are bacteria claustrophobic? The problem of micrometric spatial confinement for the culture of micro-organisms

Céline Molinaro, *[a] Violette Da Cunha, [b] Aurore Gorlas, [b] François Iv, [a] Laurent Gallais, [a] Ryan Catchpole, [b] Patrick Forterre, [b] Guillaume Baffou *[a]

Culturing cells confined in microscale geometries has been reported in many studies this last decade, in particular following the development of microfluidic-based applications and lab-on-a-chip devices. Such studies usually examine growth of *Escherichia coli*. In this article, we show that *E. coli* may be a poor model and that spatial confinement can severely prevent the growth of many micro-organisms. By studying different bacteria and confinement geometries, we determine that the growth inhibition observed for some bacteria results from fast dioxygen depletion, inherent to spatial confinement, and not to any depletion of nutriments. This article unravels the physical origin of confinement problems in cell culture, highlighting the importance of oxygen depletion, and paves the way for the effective culture of bacteria in confined geometries by demonstrating enhanced cell growth in confined geometries in the proximity of air bubbles.

## Introduction

The efficiency of cell culture *in vitro* not only depends on the selection of a proper culture medium and its chemical properties, but also depends upon *physical* parameters, such as temperature, surface composition (coating, electrical charge, …), and pressure. All these parameters have been widely investigated, well understood and generally well controlled experimentally. However, a physical parameter that has been less well-considered is spatial confinement. For some experimental methodologies living cells need to be spatially confined over micrometer scales, such as sandwiched between two coverslips, passing through constrictions, or incorporated within a microfluidic channel.[1,2] Microfluidic devices in particular have been increasingly used in this last decade to study micro-organisms and address several biological questions, from the growth and motility of bacteria, to chemotaxis, and intercellular interactions.[3,4] Additionally, microfluidics have enabled the observation of biofilm formation,[5] conjugation between strains,[6] and new insights into bacteria-mediated cancer therapies.[7] A valuable feature of microfluidic devices is the possibility to create local chemical gradients, facilitating studies of bacterial chemotaxis.[8] Furthermore, spatial constrictions like micropores can enable the selective passage of bacteria, or the observation of their behavior within a maze.[9,10] Very thin, elongated constrictions can also be used to spatially organize bacteria over the field of view of a microscope and achieve easy statistic characterization of a large number of cells.[11]

It is notable that most of these studies utilizing spatial confinements were carried out with *Escherichia coli* bacteria as a model species, which is indeed recognized as suited for culture in confined geometries. For instance, the growth rate of *E. coli* was shown to be equivalent in micron-sized channels and in suspension,[12] with a no-growth threshold around 0.5 μm.[13] Mannick et al.[14] compared the growth and motility of *E. coli* (Gram negative) and *Bacillus subtilis* (Gram-positive) in microfabricated channels, showing that *E. coli* can grow in narrower channels than *Bacillus subtilis* thanks to its thin Gram-negative cell wall. Only a handful of other prokaryotic species have been investigated in similar confinements. For instance, *Halobacterium salinarium*, an archaeon, can grow in 1.3 μm high and 10 μm long cavities, albeit with a generation time three times longer than in suspension.[15] We speculate that the predominance of *E. coli*-based literature involving microfluidic channels may come from the inability of many micro-organisms to grow in reduced spaces, though this is not stated in the literature.

In this article, we evidence the inability of different bacteria to grow in confined geometries and unravel the origin of this growth inhibition. We show that *Escherichia coli* and *Lactobacillus reuteri* can easily divide in confined geometries (<20 μm), while *Thermus thermophilus* and *Geobacillus stearothermophilus* are unable to grow below a spatial confinement of around 300 μm. These results indicate that this claustrophobic behavior results from depletion of dioxygen in small volumes, not from nutriment deficiencies, as confinement only affected the growth of aerobic bacteria. This hypothesis is further supported by the ability of the aerobic bacteria to grow

[a.] Institut Fresnel, CNRS, Aix Marseille University, Centrale Marseille, Marseille, France. Email : celine.molinaro@fresnel.fr, guillaume.baffou@fresnel.fr
[b.] Université Paris-Saclay, CEA, CNRS, Institute for Integrative Biology of the Cell (I2BC), 91198, Gif-sur-Yvette, France.
† celine.molinaro@fresnel.fr
* guillaume.baffou@fresnel.fr

in confined geometries, provided an air bubble is present nearby.

## Results and discussion

During microscopic culturing experiments, we noted a striking inhibition of growth of several bacterial species when deposited between glass coverslips, typically separated by <100 µm. To demonstrate the issue and uncover its causal origin, we conducted experiments on 4 species of bacteria (listed in Table 1) in 3 geometries. These bacteria were selected for their short generation time (<45 min) and for their oxygen requirement. The three types of geometries correspond to the three following sections.

**Table 1**: Bacterial species used in this work, along with their aero/anaerobic property and generation time.

| Bacterial species | Metabolic description | Generation time | Gram stain |
|---|---|---|---|
| *Escherichia coli* | Facultative anaerobe | 15-20 min | Gram-negative |
| *Lactobacillus reuteri* | Facultative anaerobe | ~ 45 min | Gram-positive |
| *Thermus thermophilus* | Strictly aerobe | 20 min | Gram-negative |
| *Geobacillus stearothermophilus* | Strictly aerobe | 25 min | Gram-positive |

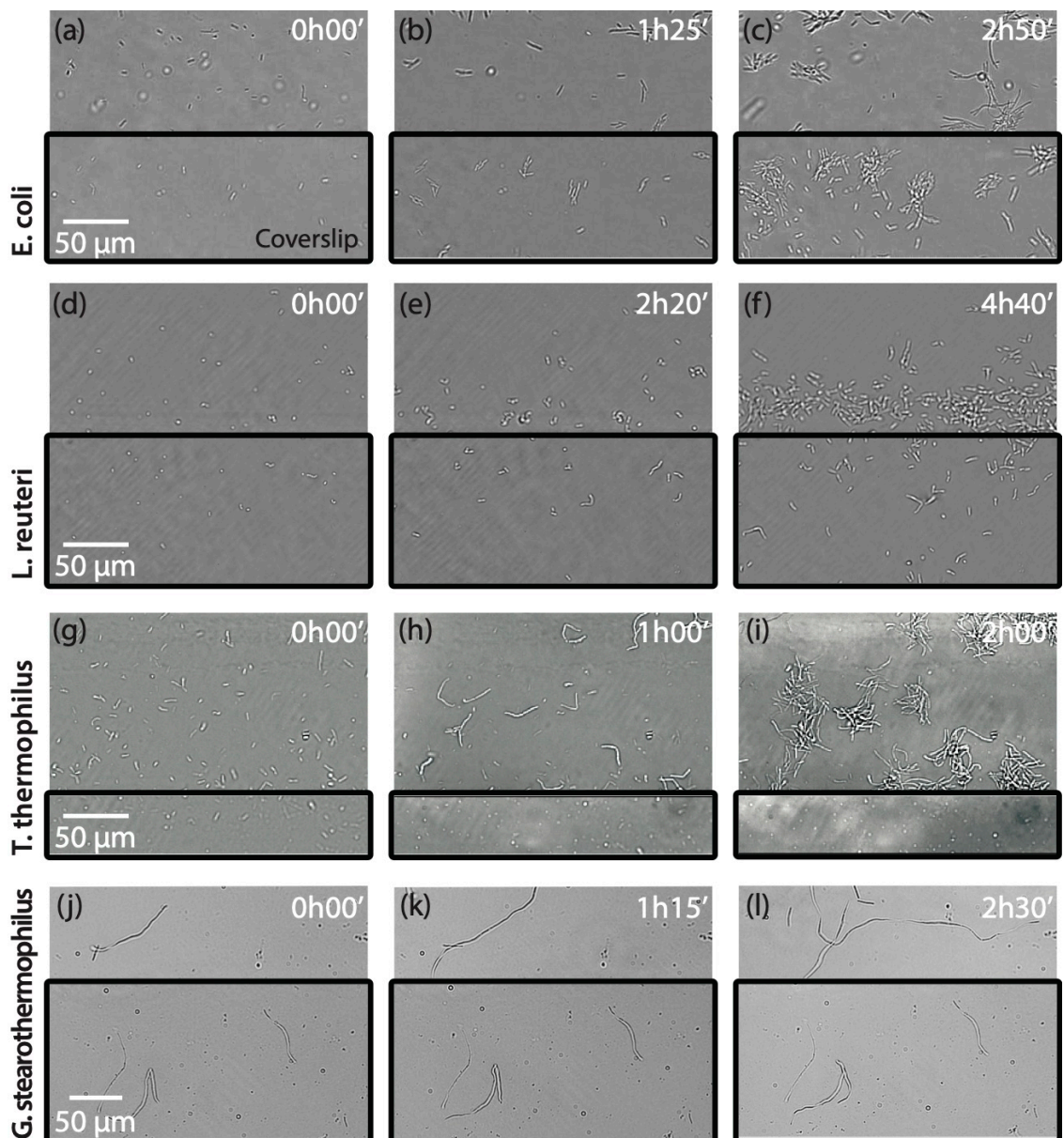

**Fig. 1**: (a, b, c) Images of the growth of *E. coli* at 37°C, at 0h, 1h25 and 2h50. (d, e, f) Images of the growth of *Lactobacillus reuteri* at 35°C, at 0h, 2h20 and 4h40. (g, h, i) Images of the growth of *Geobacillus stearothermophilus* at 60°C, at 0h, 1h and 2h. (j, k, l) Images of the growth of *Thermus thermophilus* at 70°C, at 0h, 1h15 and 2h30. In all the images, the top coverslip lies on the bottom part, represented as a thick, black solid line.

**Growing bacteria in < 20-μm thin layer.**

We first demonstrate the ability/inability of bacteria to grow in confined geometries. Bacterial species depicted in Table 1 were cultured separately, sandwiched between two coverslips, in a 20-μm thick liquid medium. The top coverslip was cut to be smaller than the bottom, so that only part of each culture was covering. As such, some bacteria of a given culture were spatially confined, while others were not. Videos were acquired over several hours at the edge of the coverslip, to capture both confined and unconfined bacteria within a single field of view of the microscope (Figure 1).

Under these conditions, *E. coli* grew normally both under the coverslip and in open space with a similar generation time (see Movie Ecoli.mov in Suppl. Info., from which images in Figs. 1a-c have been extracted). This result is consistent with the many effective studies reported on *E. coli* bacteria in confined geometries [11–13] (as discussed in above). We reproduced the same experiments with *L. reuteri* (Figs. 1d-f), which also displayed growth both under the coverslip or in open space with a similar generation time (see Movie LactobacillusR.mov in Suppl. Info.). In contrast, results obtained with the aerobic bacteria *G. stearothermophilus* (Figs. 1g-i) revealed a completely different behavior (see Movie GeobacillusS.mov in Suppl. Info.) - substantial growth was observed on the open side, with complete growth inhibition in the 20-μm deep medium layer. Furthermore, spatially confined *G. stearothermophilus* adopt a spherical geometry reminiscent of sporulation. Similarly, *T. thermophilus*, another strictly aerobe, exhibited the same behavior: the total inability to grow in confined geometry (Figs. 1j-l). *T. thermophilus* grows as long filaments observable in the uncovered portion of the cultured slides (Figs. 1k,l). In contrast, similar filamentous cells do not develop when covered (see the associated Movie Thermus-T.mov in Suppl. Info.).

**Claustrophobic behavior as a function of the media layer thickness.**

The previous section demonstrates the inability of some bacteria species, *G. stearothermophilus* and *T. thermophilus*, to grow in a 20 µm thick medium, while they can grow in open medium. The natural question is to determine the maximum degree of spatial confinement to which this problem persists. For the two species of bacteria that suffer when confined, *G. stearothermophilus* and *T. thermophilus*, samples were prepared with liquid thicknesses of 120 µm, 360 µm, 600 µm, 840 µm and 1 mm (by stacking 120 µm spacers, see Methods section). Each sample contained the initial same amount of bacteria (equal optical density (OD) and solution volume).

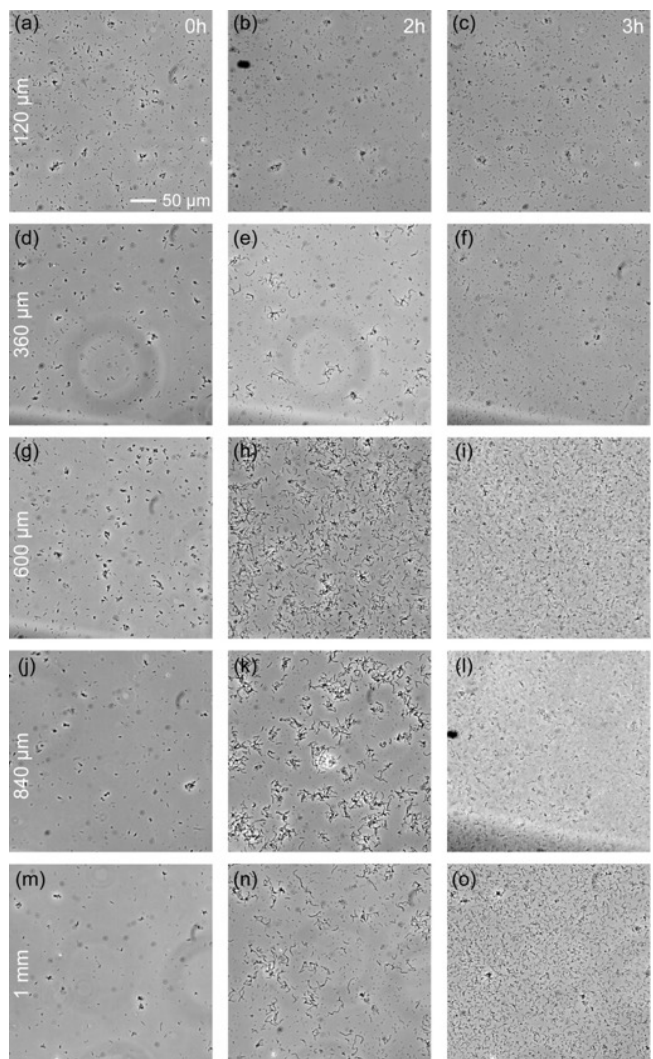

**Fig. 2**: Phase-contrast images of *Geobacillus stearothermophilus* (a-e) in various liquid thickness, from 120 μm to 1 mm (rows), after 0h, 2h and 3h (columns) of incubation at 60°C, over the same area.

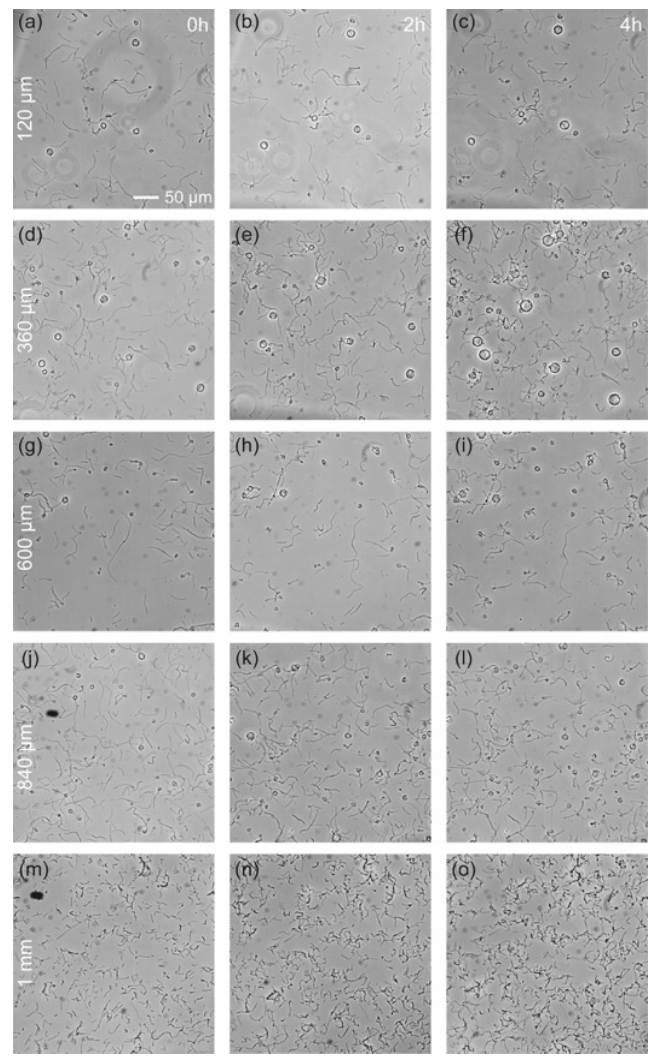

**Fig. 3**: Phase-contrast images of *Thermus thermophilus* for various liquid thickness, from 120 μm to 1 mm (a,b,c) to (m,n,o), after 0h, 2h and 4h (columns) of incubation at 70°C, over the same area.

The results obtained with *G. stearothermophilus* are displayed in Fig2. Similar to that observed above (Fig. 1h,i), a sporulation-like phenotype was observed after 2 h of incubation at 120-µm and 360-µm liquid thickness. Sporulation typically indicates unfavourable growth conditions, suggesting spatial confinement is poorly tolerated by this species. In contrast, above 360µm effective growth was observed at both 2 h and 3 h. A similar behaviour was observed for *T. thermophilus*, that is no growth below 360 µm of medium thickness, and effective growth at 360 µm and above (Fig. 3). *T. thermophilus* bacteria required more time to demonstrate effective growth, around 4h of incubation. In both cases, thicker liquid medium facilitated faster growth of bacterial cells.

**Growth close to air bubbles despite of confinement.**

We demonstrated above that only the two facultative anaerobes we studied could blossom in confined environments. These results suggest growth inhibition may be cause by a lack of dioxygen in confined environments. This hypothesis was confirmed here by the observation that effective growth of *G. stearothermophilus* and *T. thermophilus* was possible in the presence of nearby trapped air bubbles, despite confinement in 120-µm thick culture medium (Figs. 4a-f). By underfilling the sample well, an air/medium interface can be trapped within the coverslip-confined region and observed within the microscope field. The presence of this interface allowed growth of *G. stearothermophilus* in the vicinity, and even several 100 µm away. Despite the 120-µm deep culture medium, this growth was even faster than the growth observed in the 1 mm confinement tests (Figs. 4a-c vs 2m-o). Additionally, the sporulation-like phenotype was absent close to the air layer (Fig. 4b vs 1i).

Similar behaviour was observed with *T. thermophilus* in a 120-µm thick culture medium, this time illustrated with a trapped air bubble (Figure 4d-f). Here again, effective bacterial growth is observed in the proximity of the air bubble. Additionally, over the course of 4 hours the bubble reduced in size, consistent with oxygen consumption by the bacteria. For both experiments, images were recorded from the same sample but away from the air layer or bubble. No growth was observed for both types of bacteria anywhere else in the sample (Figure S2 and S3).

Air layers or bubbles behave as constant, long sources of dioxygen, allowing continued diffusion from the gas to liquid. Gas content is the only medium property expected to vary at the vicinity of an air interface, which confirms the hypothesis of dioxygen depletion, and/or carbon dioxide increase, as the origin of the deleterious effect and apparent claustrophobic behaviour of aerobic bacteria: *bacteria are actually suffocating in confined geometries*.

**Estimation of oxygen consumption in confined geometries.**

We next wished to estimate the time needed for the bacteria to consume the dioxygen contained in the closed volume in which they are living. Such a calculation can be compared to the results presented above and verify the consistency of an oxygen depletion mechanism. To the best of our knowledge, no measurement of dioxygen consumption of *G. stearothermophilus* or *T. thermophilus* have been reported in the literature. Typical values of dioxygen molecule consumption rate per bacteria $Q_0$ found in the literature for other bacteria[16,17] including *E. coli*, lie in the range of $Q_0$= 3×10$^{-19}$ to 3×10$^{-18}$ mol/s/cell, representing 2×10$^5$ to 2×10$^6$ molecules/s/cell. In the experiment reported above, for the case of a 120 µm spacer, we used $V$=7 µL of bacterial suspension with an OD of 0.08, corresponding to $n_b$=2×10$^8$ cell/mL. In 7 µL of water at 25°C, with $\rho_{O_2}$=8.28 mg/L of dissolved dioxygen (at 60°C, only 4.6 mg/L), there is $n_0$=1.8x10$^{-9}$ moles of O$_2$. All the dissolved dioxygen is to be consumed over a time scale $\tau = \rho_{O_2}/Q_0 M_{O_2} n_b$, where $M_{O_2}$ is the molar mass of O$_2$. Giving the typical range of O$_2$ consumption rate $Q_0$ given above, it yields a time scale $\tau$ ranging from 6 minutes to 1 hour. This range means that after a few minutes or tens of minutes, the oxygen content of the sample should be significantly reduced, hampering the proper development of strictly aerobic bacteria. This estimation is consistent with the experiments we conducted.

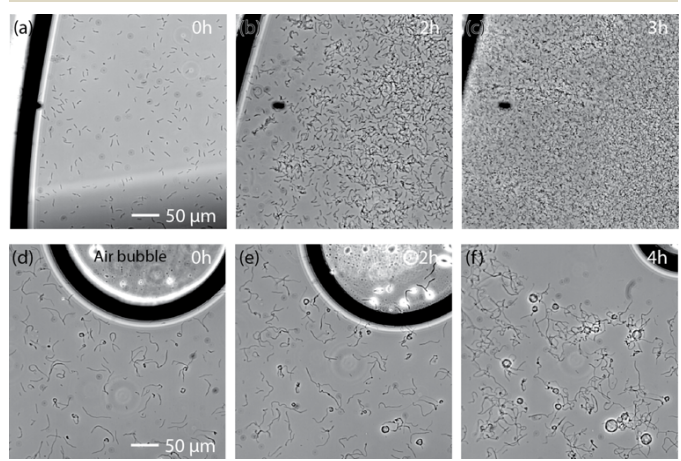

**Fig. 4**: Phase-contrast images of *G. stearothermophilus* after (a) 0h, (b) 2h and (c) 3h of incubation at 60°C. All the images correspond to the same location within the sample. The air layer is on the left side of each image demarcated by the thick black line. Phase-contrast images of *T. thermophilus* after (d) 0h, (e) 2h and (f) 4h of incubation at 70°C. Images display the same location within the sample. An air bubble, located at the top-right corner of each image, was trapped in the suspension, shrinking over time.

## Conclusions

We demonstrate that spatial confinement below a couple of hundred micrometers, prevents the growth of aerobic bacteria. By measuring the growth of 4 bacterial species (two (facultative) anaerobes and two aerobes) under such conditions, we demonstrate that this phenomenon is unique to the strictly aerobic species. We hypothesized that this growth inhibition was due to fast oxygen consumption and depletion. In support of this hypothesis, we show that growth of aerobic cells can be restored in the vicinity of trapped air bubbles, even under confinement conditions. Not only do these results provide evidence toward the origin of the problematic growth we observed (oxygen depletion), it also provides a convenient approach to enable aerobic bacteria culture under confined geometries, such as microfluidic devices. We recognize that the origin of the claustrophobic behavior may not be singular. In particular, a possible explanation could have been rapid depletion of a nutriment present in the medium due to consumption by the cells, or accumulation of inhibitory waste compounds. Additionally, this effect could be synergistic with rapid hypoxia resulting in accumulation of by-products from anaerobic metabolism, or depletion of nutriments preferred in hypoxic conditions. Also, if some nutriments had electrostatic affinity with the naturally negatively charged glass surface, it could lead to an effective removal of these compounds from the culture medium bulk, or even pH modification. This mechanism would be favored by the presence of the two glass coverslips and the sandwich geometry, characterized by a very high glass-surface/medium-volume ratio, which does not occur with a single coverslip. The results presented in this study challenge any such variations of the chemical nature of the culture medium as a possible explanation of the problem. A lack of dioxygen was evidenced as the sole origin of the problem. *Bacteria are actually suffocating in confined geometries.*

## Methods and Materials

**Bacteria growth conditions.**
*Escherichia coli* (HST08 – Stellar) were grown at 37°C, 200 rpm in LB media (LB Broth 1231 - Conda). *Lactobacillus reuteri* (DSM 20016) were grown at 35°, 200 rpm in MRS culture media (MMRS broth 69966 Sigma-Aldrich and tween 80 P8074 Sigma-Aldrich). *Thermus thermophilus* (CIP 110185T, type strain HB8) were grown at 70°C, 200 rpm in LB media. *Geobacillus stearothermophilus* were grown at 60°C, 200 rpm in LB media. For each experiment, cultures were grown overnight. The optical density (OD) was then measured with a
spectrophotometer (Ultrospec 10 – Biochrom) and adjusted to the required OD (600nm).

**20 µm confinement experiments.**

A VAHeat heating stage (Linnowave) with PDMS (Polydimethylsiloxane) reservoirs (Linnowave) was used to set the temperature of the culture (Fig S1). Observation were performed with a 40× air-objective (Olympus). To avoid excess convection and heat shock, heating was performed gradually (~0.05°C/s). Experiments were conducted at least in triplicate for each bacterial strain. In each experiment, 0.2 µL of bacterial suspension (OD 0.3) was dropped in the VAHeat reservoir and immediately covered by a small 150-µm thick coverslip with a triangular shape. This top coverslip was fabricated by a glass laser cutting technique. The laser processing system is based on a commercial femtosecond-diode-pumped ytterbium amplified laser source (S-Pulse HP, Amplitude Systemes) operating at 1030 nm (FWHM 5 nm) with a spatially Gaussian beam profile. The source is coupled in a dual-axis scanning galvometric system (GVS012/M, Thorlabs) with metallic mirrors and a 100 mm focal length f-theta lens (FTH100-1064, Thorlabs). The 150 µm thickness coverslips were processed with the following parameters: beam diameter 60 µm, repetition rate 1 kHz, 0.5mJ energy per pulse, 10µm steps.

**Varying thicknesses experiments.**

To create cavities of various thicknesses, we used spacers, 1 mm thick (Press-to-seal Silicone Isolator, P24744, Invitrogen) and 120-µm thick (SecureSeal Imaging spacer, GBL654008, Sigma-Aldrich). The 120-µm spacers have been stacked to gradually vary the thickness from 120 µm to 840 µm. The same volume (7 µL - OD 0.08) of bacterial suspension was deposited in each seal. Each well was then completely filled with the appropriate culture media, and hermetically sealed with a glass coverslip. After a sedimentation time (from 1h30 to 20h depending on the strain and on the well thickness), the samples were incubated at the appropriate growth temperature without shaking. Images of the sample were taken just before incubation (0h) and then every hour with a CMOS camera (DCC1545M-GL - Thorlabs) plugged to an inverted phase-contrast microscope (Eclipse TS2 – Nikon, 10× objective). A mark was placed on each sample to observe the same area throughout the experiment. Only *Geobacillus stearothermophilus* and *Thermus thermophilus* were studied this way.

**Bubble experiments.**

The growth of bacteria was monitored in the vicinity of air in order to validate the hypothesis of claustrophobia due to weak dioxygen renewal. 120 µm thick spacer were stuck on cleaned glass coverslides. 5 µL of bacterial suspension (OD 0.08) was deposited in each seal. Then, the 120 µm thick spacer was enclosed with a glass coverslip. The seals were not complete, with a thin layer of air between the spacer and the suspension. Air bubbles trapped randomly inside the suspension droplet were observed (Figure 7). After waiting 1h30 for sedimentation, samples were incubated at 60°C for *Geobacillus stearothermophilus* and 70°C for *Thermus thermophilus*. Images of the sample were taken just before incubation (0h) and then every hour with a CMOS camera plugged to an inverted phase-contrast microscope. Experiments were performed at least in triplicates.

## Conflicts of interest

There are no conflicts to declare

## Acknowledgements

This project has received funding from the European Research Council (ERC) under the European Union's Horizon 2020 Research and Innovation Programme (grant agreement no. 772725, project HiPhore).